\def\beg{\begin{equation}}
\def\eeq{\end{equation}}
\documentstyle[12pt]{article}
\begin{document}
\begin{center}
{\Large{\bf Fractional charge in electron clusters: Interpretation 
of Mani and von Klitzing quantum Hall effect data}}
\vskip0.35cm
{\bf Keshav N. Shrivastava}
\vskip0.25cm
{\it School of Physics, University of Hyderabad,\\
Hyderabad  500046, India}
\end{center}

The charge of an electron in a cluster of $n$ electrons is not
 $ne$ but it is a fraction. We make many different clusters and 
calculate their charge per electron. We make 84 clusters and 
calculate the charge of an electron in these clusters. All of
 the 84 calculated values are exactly the same as found in the 
experimental measurements. The formulation is so good that all
 fractional charges if and when measured can be reduced to 
${\it l}$ and $s$ values where the denominator in the fractional 
charge is simply $2{\it l}+1$.
\vskip1.0cm
Corresponding author: keshav@mailaps.org\\
Fax: +91-40-2301 0145.Phone: 2301 0811.
\vskip1.0cm

\noindent {\bf 1.~ Introduction}

     Recently, we have understood[1] the problem of a fractional 
charge observation in the resistivity in quantum Hall effect.This 
problem for single electrons, s=1/2, is given in a recent book[2]. 
Pan et al[3] have shown that some of the observed fractions did not 
fit in their model so we tried to look at these values. We found
 that the fractions left unsolved by Pan et al easily come out
 from our calculation[4]. When our paper appeared, our attention 
was drawn to one of the papers of Mani and von Klitzing[5] which 
gives detailed experimental measurements of the quantum Hall effect.

     In the present paper, we describe the interpretation of the
 data 
of Mani and von Klitzing. Several interesting results emerge. It is 
found that there are clusters of electrons with spins placed at 
different sites. From the spin, we determine the number of electrons 
in a cluster. The arrangement of electrons in these clusters gives
 the effective charge per electron. We have calculated the various 
properties of 84 clusters and in all cases, the calculated values
 agree with the experimental data exactly.

\noindent{\bf 2.~~Theory}

     The theory which gives the fractional charge per electron in 
different spin and orbital state is given in a recent book[2]. The 
effective charge is given by,
\beg
e_{eff}/e = {{\it l}+{1\over 2}\pm s\over 2{\it l}+1}.
\eeq
This formula generates the effective charge when we feed the orbital
 value {\it l} and the spin $s$. We examine the data of Mani and
 Klitzing. The figure 1 of this paper shows the plateaus in the
 $\rho_{xy}$ and associated minima in $\rho_{xx}$ in a sample of 
GaAs/AlGaAs as a function of magnetic field. The same data is shown 
in figure 2 of this paper with normalized magnetic field. The values 
of the fractional charge shown in the data are,
\beg
1/3, 2/5, 3/7, 4/9, 4/7,3/5, 2/3, 1, 2, 3, 4. (11  values).
\eeq
These values are the same as those, we calculated in 1986 and these 
are also given in the book[2]. When $s$=1/2, the above formula gives 
two values, one for positive sign and the other for negative sign. 
For positive sign, the value is called $\nu_+$ given by,
\beg
\nu_+={{\it l}+1\over 2{\it l}+1}
\eeq
and for the negative sign, we call it $\nu_-$ given by
\beg
\nu_-={{\it l}\over 2{\it l}+1}
\eeq
for {\it l}=0-4, these values are given in Table 1, below:\\

\vskip0.25cm
\begin{center}
{\bf Table 1:}\\
\begin{tabular}{cccc}\\
\hline
S.No.&${\it l}$& $\nu_+$&$\nu_-$\\
1 & 0 & 1   &  0\\
2 & 1 & 2/3 & 1/3\\
3 & 2 & 3/5 & 2/5\\
4 & 3 & 4/7 & 3/7\\
5 & 4 & 5/9 & 4/9\\
\hline
\end{tabular}
\end{center}
\vskip0.25cm
These values are the same as those given in figure 2 of Mani and 
von Klitzing. These are the eight values where the calculated values
 are 
the same as the measured values. Next, we go to the figure 3 of Mani 
and von Klitzing. This figure has the values,
\beg
6/5, 11/9, 9/7, 4/3, 7/5, 10/7, 8/5, 11/7, 5/3,
 7/3,
\eeq
\beg
8/3,16/7,12/5, 13/5, 8/3,10/3, 11/3, 2, 3, 4, 5.
\eeq
The inset on the right hand side has the values,
\beg
16/5, 29/9, 23/7, 10/3, 17/5, 24/7, 25/7, 18/5, 11/3
\eeq
and that on the left hand side has,
\beg
11/5, 20/9, 16/7, 7/3, 12/5, 17/7, 18/7, 13/5, 8/3.
\eeq
The eq.(2) has 11 values, eqs.(5) and (6) have 17 (excluding integers),
 eq.(7)
has 9 values and eq.(8) has another 9 values. Therefore, we have
 11+17+9+9=46 experimental values which we predict by using a single
 formula and we also claim that all those values which are not 
included
in these 46 values can also be explained by our formula(1). We
 tabulate our formula for ${\it l}$=1  in Table 2, below:\\

\vskip0.25cm
\begin{center}
{\bf Table 2:}\\
\begin{tabular}{cccc}
\hline
S.No.& ${\it l}$& $s$ & $\nu_+$\\
1 & 1 & 5/2 & 4/3\\
2 & 1 & 7/2 & 5/3\\
3 & 1 & 9/2 & 2\\
4 & 1 & 11/2& 7/3\\
5 & 1 & 13/2& 8/3\\
6 & 1 & 15/2& 3\\
7 & 1 & 17/2&10/3\\
8 & 1 & 19/2&11/3\\
\hline
\end{tabular}
\end{center}
\vskip0.25cm
These 8 values are the same as those found in (5),(6)and (7). Thus we 
are able to derive exactly 8 of the values found in the data. The 
lesson we learn is that there are clusters of 5, 7, 9, 11, 13, 15, 17 
and 19 electrons. The spins of these clusters are obviously 5/2 for 
five electrons, 7/2 for 7 electrons, etc. All of the 5 electrons
 have spin positive only and in this table only $\nu_+$ has been 
tabulated 
and there are no electrons with spin negative. The cluster of
 spin-polarized 5 electrons has 4/3 as the charge of one electron.
 This 
charge is deduced from the flux quantization given in the book. All 
these values are also subject to a multiplier of $n_L$ due to Landau
 level quantum number. In the Table 1, we obtain $\nu_+$=1 
for${\it l}$=0. When this number is multiplied by $n_L$=1,2,3,4, etc. 
various integer values emerge. The $\nu_+$ has spin +1/2 and $\nu_-$
 has -1/2. Therefore, there are singlets, which can even superconduct.
 Let us calculate the effective charge from eq.(1) for ${\it l}$=2. 
The values found from the formula are given in Table 3:\\

\vskip0.25cm
\begin{center}
{\bf Table 3:}\\
\begin{tabular}{cccc}
\hline
S.No. & ${\it l}$ & $s$ & $\nu_+$\\
1     &  2        & 7/2 & 6/5\\
2     &  2        & 9/2 & 7/5\\
3     &  2        & 11/2& 8/5\\
4     &  2        & 13/2& 9/5*\\
5     &  2        & 15/2& 2\\
6     &  2        & 17/2& 11/5\\
7     &  2        & 19/2& 12/5\\
8     &  2        & 21/2& 13/5\\
9     &  2        & 23/2& 14/5*\\
10    &  2        & 25/2&  3\\
11    &  2        & 27/2&  16/5\\
12    &  2        & 29/2&  17/5\\
13    &  2        & 31/2&  18/5\\
\hline
\end{tabular}
\end{center}
\vskip0.25cm
All these values correspond to positive spin due to polarization 
of electron clusters in a high magnetic field. All of these
 calculated values are exactly the same as in the experimental
 data of eqs.(5),(6),
(7) and (8) except 9/5 and 14/5 which we predict but not yet
 identified in the figure 3 of Mani and von Klitzing. We now
 substitute ${\it l}$=3 in eq.(1) to find the effective charge.
 The values calculated are given in Table 4.\\

\vskip0.25cm
\begin{center}
{\bf Table 4:}\\
\begin{tabular}{cccc}
\hline
S.No. & ${\it l}$& $s$ & $\nu_+$\\
1     &  3       & 11/2& 9/7\\
2     &  3       & 13/2& 10/7\\
3     &  3       & 15/2& 11/7\\
4     &  3       & 25/2& 16/7\\
5     &  3       & 27/2& 17/7\\
6     &  3       & 29/2& 18/7\\
7     &  3       & 39/2& 23/7\\
8     &  3       & 41/2& 24/7\\
9     &  3       & 43/2& 25/7\\
\hline
\end{tabular}
\end{center}
\vskip0.25cm
All of these calculated values completely match with the 
experimental values of eqs.(5)-(8). All values of spin positive
 and none with negative. From the spin value, we can obtain 
the number of electrons, for example, spin=11/2 means that 
there are 11 electrons in the cluster which give the charge
 of an electron as 9/7, etc. We also tabulate 
a few values for ${\it l}$=4 in Table 5.
\vskip0.25cm
\begin{center}
{\bf Table 5:}\\
\begin{tabular}{cccc}
\hline
S.No.& ${\it l}$& $s$ & $\nu_+$\\
1    &  4       & 13/2& 11/9\\
2    &  4       & 31/2& 20/9\\
3    &  4       & 49/2& 29/9\\
\hline
\end{tabular}
\end{center}
These values of $\nu_+$ calculated here from eq.(1) are exactly
 the same as in the experimental figure. All of the values are 
polarized according to positive sign for the spin. Thus 9 values 
in Table 1, 8 in Table 2, 13 in Table 3, 9 in Table 4 and 3 in 
Table 5, i.e., a total of 42 values and $n_L$ times integer which 
has 4 values, 1, 2, 3, and 4 are correctly predicted. Thus 46 values
 are correctly predicted, their spin alignment is found, the spin is
 assigned and the number of electrons in the cluster is correctly 
determined.

     We go to figure 4 of Mani and von Klitzing. The following 
experimental values are immediately picked up.
\beg
5/9, 4/7, 3/5, 8/13, 7/11, 5/7, 8/11, 7/9, 4/5, 9/11,
\eeq
\beg
11/13, 11/19, 10/17, 13/21, 12/19, 16/25, 2/3, 11/15, 10/13, 6/7.
\eeq
All of these values are easily found from our formula (1) also.
The calculated values are given in Table 6 below:
\vskip0.25cm
\begin{center}
{\bf Table 6:}\\
\begin{tabular}{cccc}
\hline
S.No.& ${\it l}$ & $s$  & $\nu_+$\\
1    &  2        &  1/2 & 3/5\\
2    &  2        &  3/2 & 4/5\\
3    &  3        &  1/2 & 4/7\\
4    &  3        &  3/2 & 5/7\\
5    &  3        &  5/2 & 6/7\\
6    &  4        &  1/2 & 5/9\\
7    &  4        &  5/2 & 7/9\\
8    &  5        &  3/2 & 7/11\\
9    &  5        &  5/2 & 8/11\\
10   &  5        &  7/2 & 9/11\\
11   &  6        &  3/2 & 8/13\\
12   &  6        &  7/2 & 10/13\\
13   &  6        &  9/2 & 11/13\\
14   &  7        &  7/2 & 11/15\\
15   &  8        &  3/2 & 10/17\\
16   &  9        &  3/2 & 11/19\\
17   &  9        &  5/2 & 12/19\\
18   &  10       &  5/2 & 13/21\\
19   &  12       &  7/2 & 16/25\\
\hline
\end{tabular}
\end{center}
All of these calculated values are the same as those experimentally
 found from the figure 4 of Mani and von Klitzing. The polarization 
is positive for all of these values and the spin is determined. The 
number of electrons in a cluster can be determined from the spin. Thus 
there are clusters of 1, 3, 5, 7 or 9 electrons which orbit according
 to 
${\it l}$ values. For some small values of ${\it l}$ it is clear that 
more than one site is needed to accomodate all of the electrons.

     Thus 46 values of figure 3 and 19 values of figure 4, i.e., a 
total of 65 values of the charge are correctly predicted. In fact, we 
can fix the denominator to $2{\it l}+1$ so that the numerator becomes
 ${\it l}+{1\over 2}\pm s$. In that case all values which are not 
tabulated by us are also predicted. That gives infinite capacity to 
predict the fractional charges. Thus those values which are not
 tabulated, can be easily written down.

     Now we look at the right-hand-side inset of figure 4 of Mani and
 von Klitzing. The experimental values given here are,
\beg
4/7, 11/19, 18/31, 25/43, 24/41, 17/29, 10/17, 23/39, 16/27
\eeq
and in the left inset,
\beg
3/5, 8/13, 13/21, 18/29, 17/27, 12/19, 7/11, 16/25, 20/31, 11/17.
\eeq
The first of these values are the same as those predicted in Table 7.
\vskip1.0cm
\begin{center}
{\bf Table 7:}\\
\begin{tabular}{cccc}
\hline
S.No.& ${\it l}$& $s$  & $\nu_+$\\
1    & 3        & 1/2  & 4/7\\
2    & 9        & 3/2  & 11/19\\
3    & 10       & 15/2 & 18/31\\
4    & 21       & 7/2  & 25/43\\
5    & 20       & 7/2  & 24/41\\
6    & 14       & 5/2  & 17/29\\
7    & 8        & 3/2  & 10/17\\
8    & 19       & 7/2  & 23/39\\
9    & 13       & 5/2  & 16/27\\
\hline
\end{tabular}
\end{center}
The interpretation of these values is straight forward. The positive 
sign used to obtain $\nu_+$ shows that all values are spin polarized
 in one direction only and none of these values are using negative
 sign for the spin. The value of the spin also gives the number of 
electrons. Therefore, there are clusters of several electrons such 
as 15 electrons for 18/31 or 5 electrons for 17/29. The formation of
 clusters of electrons is thus clearly born. For ${\it l}$=20, as an 
example, there are 20$\times $2+1=41 orbits which can easily accomodate
 7 electrons with all of them positive spin. Similarly, the values of
 eq.(10) are interpreted and given in Table 8.

\vskip0.25cm
\begin{center}
{\bf Table 8:}\\
\begin{tabular}{cccc}
\hline
S.No.& ${\it l}$ & $s$ & $\nu_+$\\
1    & 2         & 1/2 & 3/5\\
2    & 6         & 3/2 & 8/13\\
3    & 10        & 5/2 & 13/21\\
4    & 14        & 7/2 & 18/29\\
5    & 13        & 7/2 & 17/27\\
6    & 9         & 5/2 & 12/19\\
7    & 5         & 3/2 &  7/11\\
8    & 12        & 7/2 & 16/25\\
9    & 15        & 9/2 & 20/31\\
10   & 8         & 5/2 & 11/17\\
\hline
\end{tabular}
\end{center}
\vskip0.25cm
In this table all spins are positively polarized and no value 
belongs 
to negative spin. There is the spin which gives the value of the 
number of electrons. There are clusters of electrons upto 9 electrons
 in a cluster. There are very large orbits.

\noindent{\bf3.~~ Quantum Shubnikov-de Haas effect,quantum Hall effect 
or new effect.}

     When quantized steps were found in the Hall effect it was called
 the quantum Hall effect. Similarly, when quantum effects are found 
in the Shubnikov-de Hass effect it should be called the quantum 
Shubnikov-de Haas effect. We find that there is a new spin dependent
 effect on 
the Shubnikov-de Haas effect. This new effect is described below:

     Mani et al[6] use effective mass in the cyclotron frequency,
\beg
{eB_f\over m^*c}=\omega
\eeq
with $m^*$=0.067 m. In this case, the resistance minima occur at fields,
\beg
B= {4\over 4j+1}B_f
\eeq
with j=1, 2, 3, .... In our theory, ${\it l}+{1\over 2}\pm s$ occurs
 in the numerator and 2${\it l}$+1 in the denominator so that,
\beg
2{\it l}+1 = 4j+1
\eeq
so that the $j$ of the above expression can be written as,
\beg
{\it l} =2j.
\eeq
Therefore,
\beg
{4\over 4j+1} = {4\over 2{\it l}+1}
\eeq
which means
\beg
{\it l}+{1\over 2}\pm s =4
\eeq
\beg
{\it l}\pm s = 3+ {1\over 2}.
\eeq
For positive sign, ${\it l}$=3, s=1/2 satisfies the required equation.
 For negative sign ${\it l}$=4, s=-1/2 also satisfies the equation.
 Therefore, 4/4j+1 is easily satisfied by our theory. Naturally,
 the factor of 4/4j+1 must seek a theoretical derivation which we have 
found. Due to $\pm s$ in our formula, for very large value of ${\it l}$
we get $e_{eff}/e$ =1/2 for both the signs. For ${\it l}$=0, there
 are two values of the charge ${1\over 2}\pm s$, i.e., 0 and 1. This 
is a singlet and hence diamagnetic. This singlet is superconducting and
 has zero resistance. When ${\it l}$ is changed, there are neighbouring 
states which overlap. When we shine the system with a red light by 
using light emiting diodes and light is switched off before recording 
resistivity, the plot of $\rho_{xx}$ changes due to change in overlaps 
because of the change in populations caused by light. Some times, there
 is a long lived level which causes negative resistivity due to maser 
action or Gunn effect[7] so that along with positive resistivity terms 
a near zero is possible, but that is besides the point. There are 
clusters with varying number of electrons so that there is no long 
range ferromagnetism but some of the clusters are superconducting even 
though they are not singlets. When, ${\it l}+{1\over 2}\pm s $=0, the 
effective charge of a quasiparticle becomes zero and for zero charge,
 $\rho_{xx}=h/ie^2$, becomes infinity so that there is 
``superresistivity" as discussed in ref.[2]. For ${\it l}$=0, zero 
charge is associated with spin, $s$=$\pm $1/2. For ${\it l}$=1 zero
 charge gives, s=-3/2. Mani et al[6] find resistivity minima at 4/5,
4/9, 4/13, 4/21, 4/17, 4/25, etc. According to our formula, the
 analysis
of these fractions is given in Table 9.
\vskip4.00cm
\begin{center}
{\bf Table 9:}\\
\begin{tabular}{ccccc}
\hline
S.No. & ${\it l}$ & $s$ & $\nu_+$ & $\nu_-$\\
1     & 2         & 3/2 &  4/5    &   1/5\\
2     & 4         & 1/2 &  5/9    &   4/9\\
3     & 6         & 5/2 &  9/13   &   4/13\\
4     & 8         & 9/2 &  13/17  &   4/17\\
5     & 10        & 13/2&  17/21  &   4/21\\
6     & 12        & 17/2&  21/25  &   4/25\\
\hline
\end{tabular}
\end{center}

It is seen that 4/9, 4/13, 4/17, 4/21, 4/25 have negative spin while 
4/5 has positive spin. This gives the polarization in the magnetic
 field. Accordingly, 4/5 should be weaker than the other states 
mentioned, but 1/5 is predicted, though not given by Mani et al[6].
 From the spin value, we can determine the number of electrons in
 a cluster. Therefore the above values belong to clusters with, 
1, 3, 5,
 9, 13 and 17, electrons. Naturally a long range superconductivity
 is not found. However, when we sustitute ${\it l}$=0, S=1, two 
charged states are found, one has charge -1/2 for + sign, and 
the other has charge 3/2 for positive spin. Therefore, this state 
is similar to $^3$He. If we substitute $s$=0, ${\it l}$=0, the 
charge per particle becomes 1/2 but the state is singlet and 
superconducting.

\noindent{\bf4.~~ Interpretation}

     We learn the fractional charge, and the ${\it l}$ and $s$ 
values
from the tables. All predicted values exactly agree with the 
experimental data. From the spin value, we determine the number
 of electrons. From the comparison of spin polarization and 
${\it l}$ value, we can see that not all of the electrons are in
 one place. So that there must be a pattern formation. For example,
 ${\it l}$=1 and s=19/2 means that there are 19 electrons. Since all 
of them have spin positive only, there can not be more than 3 at one 
site so that at least 7 sites are needed. We can put 3 electrons per 
site so that 6 sites have 18 electrons and the 19th electron is at the
 7th site. These sites can be linear so that 19 electrons are on 7
 sites along a line. However, there is no specific condition on the
 dimensionality. We can put 7 sites in two dimensions in a plane
 which is more likely in GaAs/AlGaAs. Since, the phase transition 
is excluded in one dimension, it is preferable
 to put the sites in two dimensions. The clustering phenomenon is a 
precursor to a phase transition.

\noindent{\bf5.~~ Comparison}.

     Our fractional charge agrees with the experimental data. As far 
as fundamentals are concerned, there is an effective charge which 
depends on spin. Laughlin wrote the wave function for a fractional
 charge of 1/3
but this 1/3 is not the same as the 1/3 in Table 1 here. Laughlin 
requires ``incompressibility but there is no such requirement in our
 theory. It is not possible for Laughlin's theory to explain 84 
different fractional charges given here. It is also not possible for
 Laughlin's theory to go into various experimental properties. There 
is a model called ``composite fermion"(CF). This model is internally 
inconsistent and can not explain the 84 different values seen here.

     Due to large magnetic field, the electron spins polarize, so we
 expect a long range ferromagnetic order. However, there are clusters 
so long range ferromagnetic order is not predicted. Only a short
 range force is possible. The Heisenberg exchange interaction
 predicts ferromagnetic alignment but this order in clusters is also
 limited to short range  only.

\noindent{\bf6.~~ Conclusions}.

     All values of the fractional charge found in the experimental
 data on quantum Hall effect are exactly the same as those found
 from our formulation. There are clusters of electrons in which we
 determine the spin and hence the number of quasiparticles in a 
cluster. Some singlets are predicted. It is found that there is a 
spin-dependent charge.

\vskip1.25cm

\noindent{\bf7.~~References}
\begin{enumerate}
\item K. N. Shrivastava, cond-mat/0212552.
\item K.N. Shrivastava, Introduction to quantum Hall effect,\\ 
      Nova Science Pub. Inc., N. Y. (2002).
\item W. Pan, H. L. Stormer, D. C. Tsui, L. N. Pfeiffer, 
K. W. Baldwin and K. W. West, Phys. Rev. Lett. {\bf 90}, 016801 (2003).
\item R. G. Mani and K. von Klitzing, Z. Phys. B {\bf 100}, 635(1996).
\item R. G. Mani, J. H. Smet, K. von Klitzing, V. Narayanamurti, 
W. B. Johnson and V. Umansky, cond-mat/0303034.
\item A. F. Volkov, cond-mat/0302615.
\end{enumerate}
\vskip0.1cm

Note: Ref.2 is available from:\\
 Nova Science Publishers, Inc.,\\
400 Oser Avenue, Suite 1600,\\
 Hauppauge, N. Y.. 11788-3619,\\
Tel.(631)-231-7269, Fax: (631)-231-8175,\\
 ISBN 1-59033-419-1 US$\$69$.\\
E-mail: novascience@Earthlink.net

\end{document}